\documentclass[]{aa} 
\usepackage[varg]{txfonts}

\usepackage{natbib,twoopt}
\usepackage[breaklinks]{hyperref}    
\bibpunct{(}{)}{;}{a}{}{,} 
\makeatletter
\newcommandtwoopt{\citeads}[3][][]{\href{http://adsabs.harvard.edu/abs/#3}%
    {\def\hyper@linkstart##1##2{}%
        \let\hyper@linkend\@empty\citealp[#1][#2]{#3}}}
\newcommandtwoopt{\citepads}[3][][]{\href{http://adsabs.harvard.edu/abs/#3}%
    {\def\hyper@linkstart##1##2{}%
        \let\hyper@linkend\@empty\citep[#1][#2]{#3}}}
\newcommandtwoopt{\citetads}[3][][]{\href{http://adsabs.harvard.edu/abs/#3}%
    {\def\hyper@linkstart##1##2{}%
        \let\hyper@linkend\@empty\citet[#1][#2]{#3}}}
\newcommandtwoopt{\citealtads}[3][][]{\href{http://adsabs.harvard.edu/abs/#3}%
    {\def\hyper@linkstart##1##2{}%
        \let\hyper@linkend\@empty\citealt[#1][#2]{#3}}}
\newcommandtwoopt{\citeyearads}[3][][]{\href{http://adsabs.harvard.edu/abs/#3}%
    {\def\hyper@linkstart##1##2{}%
        \let\hyper@linkend\@empty\citeyear[#1][#2]{#3}}}
\makeatother

\begin{document}

\title{Magnetic properties of a long-lived sunspot}
\subtitle{Vertical magnetic field at the umbral boundary 
    \thanks{videos associated with Fig.~\ref{fig:scr} are available at 
         \href{http://no.such.adress/exists}{http://www.aanda.org}}}
\author{
    M. Schmassmann
    \and
    R. Schlichenmaier
    \and
    N. Bello {Gonz\'alez}
   }
\institute{Kiepenheuer Institut {f\"ur} Sonnenphysik (KIS), 
    {Sch\"oneckstr.} 6, D-79104 Freiburg i.Br., Germany\\
       \email{[schmassmann;schliche;nbello]@leibniz-kis.de}
   }
\date{submitted: May 16, 2018, accepted: October 5, 2018}

\abstract
  { In a recent statistical study of sunspots in 79 active regions, the vertical magnetic field component $B_\text{ver}$ 
    averaged along the umbral boundary is found to be independent of sunspot 
    size. The authors 
    of that study conclude that the absolute value of $B_\text{ver}$ 
    at the umbral boundary is the same for all spots.}
  { We investigate the temporal evolution of 
    $B_\text{ver}$ averaged along 
    the umbral boundary of one
    long-lived sunspot during its stable phase.}
  { We analysed data from the HMI instrument on-board SDO. Contours of continuum 
    intensity at $I_\text{c}=0.5I_\text{qs}$,
    whereby $I_\text{qs}$ refers to the average over the quiet sun areas,
    are used to extract the magnetic field along the umbral boundary.
    Projection effects due to different formation heights of the 
    \ion{Fe}{I}~617.3\,nm line and continuum are taken into account.
    To avoid limb artefacts, the spot is only analysed for heliocentric angles 
    smaller than 60\degr.}
  { During the first disc passage, NOAA AR 11591, 
    $B_\text{ver}$ 
    remains constant at 1693\,G with a root-mean-square deviation of 
    15\,G, 
    whereas the 
    magnetic field strength 
    varies substantially (mean 2171\,G, 
    rms of 48\,G) 
    and shows a long term variation.
    Compensating for formation height has little influence on the mean value 
    along each contour, but reduces the variations along the contour when 
    away from disc centre, yielding a better match between the 
    contours of $B_\text{ver}=1693\,$G 
    and $I_\text{c}=0.5I_\text{qs}$.
    }
  { During the disc passage of a stable sunspot, its umbral boundary can 
        equivalently be defined by 
    using the continuum intensity $I_\text{c}$ 
    or the vertical magnetic field component $B_\text{ver}$. 
    Contours of fixed magnetic field strength 
    fail to outline the umbral boundary.}

\keywords{sunspots --
   Sun: photosphere --
   Sun: magnetic fields --
   Sun: activity
   }
   
\maketitle

\section{Introduction}
The boundary between umbra and penumbra of sunspots has long been defined in 
terms of the continuum intensity $I_\text{c}$.
This brightness 
difference is the consequence of the different magneto-convective processes running in umbrae and penumbrae.
We have evaluated magnetic quantities to identify which of them may cause  
the different behaviour on the two sides of the umbral boundary.

\citetads{2011A&A...531A.118J} investigated the properties of the magnetic 
field at umbral boundaries and noted that the vertical 
magnetic field component $|B_\text{ver}|$ changes little along the boundaries 
of the ten sunspots he analysed and could neither verify nor falsify a 
dependence of the median value along the boundary on the area of the sunspot.
The ten-spot average of the median along the boundary was 1860\,G, 
whereas the 
mean of the standard deviations along the boundary was given as 
190\,G for 
Hinode/SP data.

\citetads{2015A&A...580L...1J} extended the analysis by investigating a 
4.5h time series of a forming sunspot using GFPI/VTT data and noting an increase of 
$|B_\text{ver}|$ at the migrating umbral boundary during penumbra formation and stabilization of this value 
after completion of the formation.
Shortly thereafter, that part of the umbral boundary was observed with Hinode/SP 
and a $|B_\text{ver}|$ value of 1810\,G measured.
They propose that the umbral mode of magneto-convection prevails in areas with 
$|B_\text{ver}|>B_\text{ver}^\text{stable}$, whereas outside, the penumbral mode takes over.

Following this line of investigation, \citetads{2017A&A...597A..60J} studied a pore whose $|B_\text{ver}|$ 
remained below this critical value. They found that a developing penumbra completely cannibalized the pore, thus supporting the assertion that in umbral areas with $|B_\text{ver}|<B_\text{ver}^\text{stable}$, the penumbral mode of magneto-convection takes over the umbral mode.

\citetads{2018A&A...611L...4J} 
extends the analysis of \citeyearads{2011A&A...531A.118J} 
to 88 scans of 79 different active regions again using Hinode/SP and showed that 
the ${I_\text{c}=0.5I_\text{qs}}$ contours match mostly the 
${|B_\text{ver}|=1867\,\text{G}}$ contours. A Bayesian linear regression showed 
that a model with constant $|B_\text{ver}|$ is more likely to explain the 
data than a first or second order polynomial 
with $\null\ \log\text{area}\ \null$ as independent variable. Furthermore 
the most likely ${|B_\text{ver}|=1867\,\text{G}}$, with a $99\%$ probability 
for ${1849\,\text{G}\le |B_\text{ver}|\le 1885\,\text{G}}$.
A dependence on the solar cycle could not be verified.

These findings have led to the \emph{\mbox{Jur{\v c}{\'a}k} criterion}, an empirical 
law stating that the umbral boundary of stable sunspots can be equivalently defined by a 
continuum intensity $I_\text{c}$ or a vertical magnetic field component 
$|B_\text{ver}|$. 
In other words, in areas with $|B_\text{ver}|>B_\text{ver}^\text{stable}$, only the umbral mode of convection exists, 
hindering other modes of magneto-convection. A conjecture can also be stated from these findings: 
umbral areas with $|B_\text{ver}|<B_\text{ver}^\text{stable}$ are unstable against more vigorous modes of convection, 
that is, they are prone to vanish.

In this work we have investigated the behaviour of the magnetic field along the 
umbral boundary in a time series of a single stable sunspot.
We used the spot of NOAA AR 11591 during its first disc passage.
This allows us to verify whether $\left<B_\text{ver}\right>_\psi(t)$ remains constant over $\approx$10 days, 
which would provide support to the \emph{\mbox{Jur{\v c}{\'a}k} criterion}. 
Hereby $\langle\cdot\rangle_\psi$ stands for average along the $I_\text{c}$ contour.

\section{Data and analysis}\label{sec:methods}

The used data are retrieved after processing by the Solar Dynamics Observatory's 
(SDO) Helioseismic and Magnetic Imager (HMI) Vector Magnetic Field Pipeline 
\citepads{2014SoPh..289.3483H} 
cutout service for NOAA AR 11591. 
Using this NOAA AR number on 
\url{http://jsoc.stanford.edu/ajax/exportdata.html}
in the \verb|im_patch| processing option automatically gives the reference 
coordinates listed in the final three columns of Table~\ref{tab:time}. 
$t_\text{<E}$ and $t_\text{>W}$ are the first and last time steps processed, 
where $t_\text{<E}$ is before the sunspot rotates over the east limb onto the 
sun and $t_\text{>W}$ is after the sunspot 
rotates off the 
west limb. A cutout size of $500\times500$\, pixel was chosen.
The data series used are \verb|hmi.Ic_noLimbDark_720s| \& \verb|hmi.B_720s|. For the full disc passage, there are 1599 time steps.

For the 180\degr-disambiguation the potential acute solution provided by the 
pipeline was adopted. This can be done using 
\href{https://hesperia.gsfc.nasa.gov/ssw/sdo/hmi/idl/hmi_disambig.pro}
{\texttt{hmi\_disambig}} with \texttt{method=0}.
We note that for all pixels $180\degr$ must be added because the azimuth is defined 
relative to the positive y-axis of the maps in CCD-frame and 
\href{http://jsoc.stanford.edu/ajax/exportdata.html}{\texttt{exportdata}}'s 
\verb|im_patch| option rotates the maps $180\degr$ so that solar north is up.

The heliographic Stonyhurst coordinates are calculated using procedures 
modified from and tested against \verb|sswidl|'s \verb|wcs| routines 
\href{https://hesperia.gsfc.nasa.gov/ssw/gen/idl/wcs/fitshead2wcs.pro}
{\texttt{fitshead2wcs}}, 
\href{https://hesperia.gsfc.nasa.gov/ssw/gen/idl/wcs/wcs_get_coord.pro}
{\texttt{wcs\_get\_coord}}, 
\href{https://hesperia.gsfc.nasa.gov/ssw/gen/idl/wcs/wcs_convert_from_coord.pro}
{\texttt{wcs\_convert\_from\_coord}} and those they call
\citepads[see][]{2006A&A...449..791T}.
The canonical value for HMI of $R_\sun=696\text{Mm}$ is used.
The transformation of the magnetic field vector into the local reference frame 
was performed with a code modified from and tested against Xudong Sun's 
\verb|sswidl| routine
\href{https://hesperia.gsfc.nasa.gov/ssw/sdo/hmi/idl/hmi_b2ptr.pro}
{\texttt{hmi\_b2ptr}} (see \citealtads{1990SoPh..126...21G};
\citealtads{2006A&A...449..791T}; \citealtads{2013arXiv1309.2392S}).

\paragraph{Quiet sun intensity.}
The limb darkening correction in the HMI pipeline was based on 
\citetads[][Eq. 9]{1977SoPh...51...25P}, which does not consider all orbital 
artefacts introduced into the continuum intensity $I_\text{c}$ of SDO/HMI data. 
Even after limb darkening removal and normalization there is a change over the 
day in $I_\text{c}$ of the order of $1\%$ towards the limb with opposite signs 
on the western and eastern hemisphere.
To compensate for this, the quiet sun intensity $I_\text{qs}$ for each time step 
was chosen such that
$I_\text{qs}$ is the mean of all the quiet sun pixels within the $500\times500$ 
cutout, where quiet sun is defined as having $I_\text{c}>0.9I_\text{qs}$.

Contours were taken at $I_\text{c}=0.4\,\&\,0.5\,I_\text{qs}$, and the 
positions of the contours are used to interpolate the values of the vertical 
magnetic field component $B_\text{ver}$, the 
magnetic field strength $|B|$ and 
the inclination to the surface normal $\,\gamma_\textsc{lrf}$.
Vertical is to be understood in the local reference frame, 
in other words, it is the direction of the surface normal.

Due to different formation heights of $I_\text{c}$ and the 
\ion{Fe}{I} 617.3\,nm line, as well as the Wilson depression 
\citepads{1774RSPT...64....1W} and differential line-of-sight opacity effects 
(see e.g. \citealtads{1995A&A...298..260R}; \citealtads{2001ApJ...547.1130W}, and 
\citeyearads{2001ApJ...547.1148W}), 
the magnetic contours are projected towards the limb (i.e. outwards) 
relative to the intensity contours. To compensate for these shifts 
and get a better match between $I_c$ and $B_\text{ver}$ contours, 
we transformed the coordinates obtained 
from $I_\text{c}$ contours, $(x,y)$, using
\begin{align}
    (x',y')&=\left(1+\frac{\Delta h}{R_\sun}\right)(x,y)\label{eq:trans}
\end{align}
before retrieving the magnetic field values at coordinates $(x',y')$.
$(x,y),(x',y')$ are helio-projective coordinates in arc-seconds from disc 
centre and $\Delta h$ is the formation height difference. 
Later on, when the contours from magnetic field maps are plotted onto the $I_\text{c}$ map (cf. Sect.~\ref{subsec:contours} and 
Figs.~\ref{fig:scr}~and~\ref{fig:scr1st0.5_-60}), the inverse of Eq.~\ref{eq:trans} is 
applied, meaning that the magnetic contours are shifted inward. 
The value of $\Delta h=465\,\text{km}$ 
results from a minimization procedure, which 
is explained on page \pageref{par:whatH}.
The effect of neglecting this compensation is discussed in Sect.~\ref{sec:whyH}.

The limits of the time series we analyse are given as $t_\text{start}$ and 
$t_\text{end}$ in Table~\ref{tab:time}. A total of 1063 time steps in this time range are available.
This time range was chosen to select data sets, for which the heliocentric angle%
\footnote{%
We note the subtle difference between the heliocentric angle and the angle between the LOS and the local vertical.
The heliocentric angle, $\theta$, is the angle between the centre of the umbra and the observer as measured from 
the centre of the sun. The angle, $\alpha$, between the LOS and the local vertical at the umbral centre is given by: 
$\alpha = \theta + r$, whereby $r = \sqrt{x^2+y^2}.$ 
For any position on the solar disc, $r$ is smaller than $r_\sun\approx 0.27\degr$. The angle, $\alpha$, is used to transform between 
the LOS and LRF coordinate systems.%
} 
of the centroid of the umbra was smaller than 60\degr.%

\begin{figure*}[p] 
    \centering
    \includegraphics[width=\textwidth]{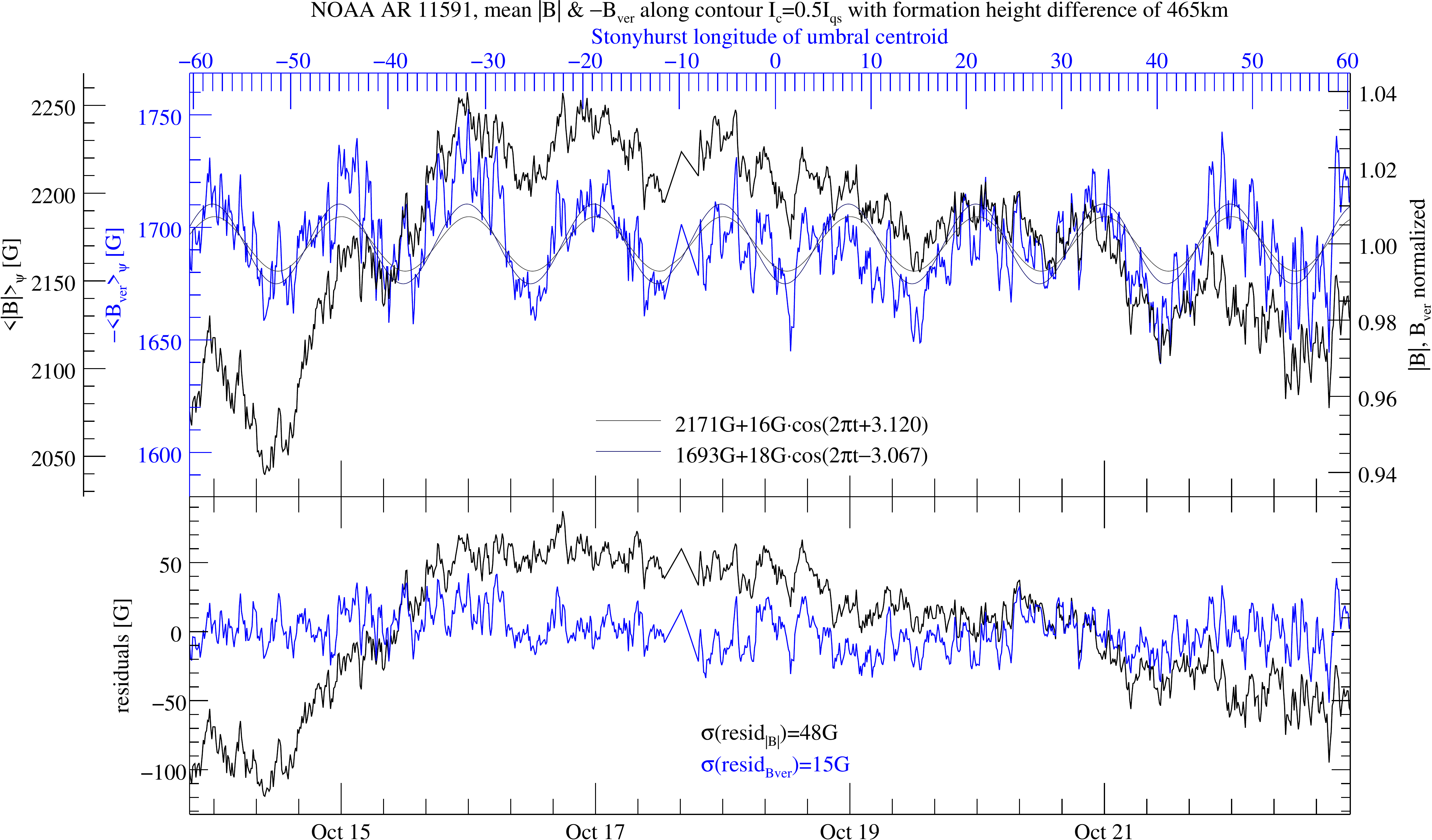}
    \caption[$\left<|B|\right>_\psi(t)$ and 
    $\left<B_\text{ver}\right>_\psi(t)$, where $I_\text{c}=0.5I_\text{qs}$ 
    and $\Delta h=465\,\text{km}$ for NOAA AR 11591]%
    {   Mean 
        magnetic field strength $\left<|B|\right>_\psi(t)$ in black 
        (it's vertical component 
        $\left<B_\text{ver}\right>_\psi(t)$ 
        in blue) along the $I_\text{c}=0.5I_\text{qs}$ contour, 
        with $\Delta h=465\,\text{km}$ accounted for, 
        Sinusodial fits and 
        the residuals for NOAA AR 11591.
    }
    \label{fig:1stH0.5}%
\end{figure*}

\begin{figure*}[p] 
    \centering
    \includegraphics[width=\textwidth]{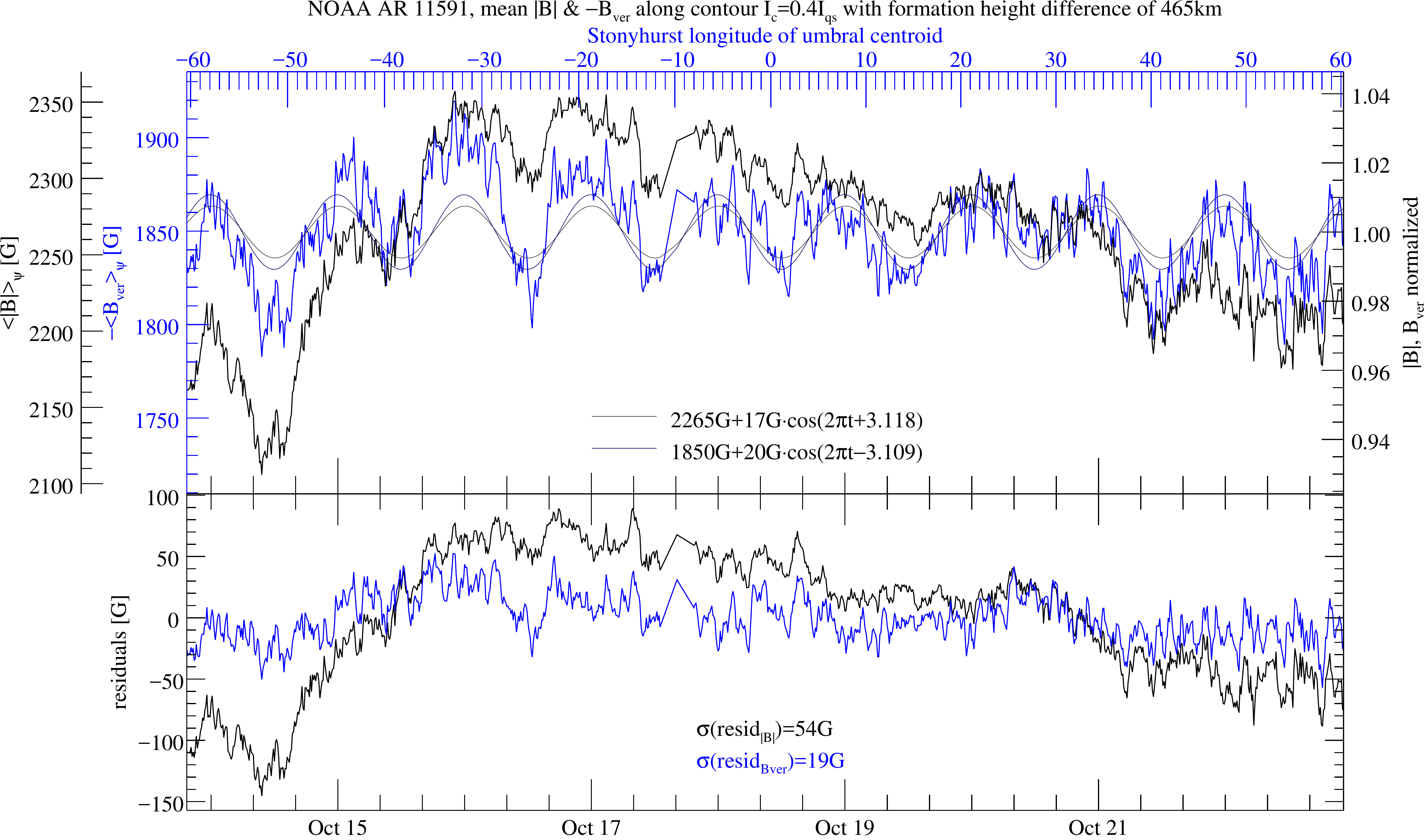}
    \caption{As Fig.~\ref{fig:1stH0.5}, but from contours at 
        $I_\text{c}=0.4I_\text{qs}$.}
    \label{fig:1stH0.4}%
\end{figure*}

\paragraph{Time series fit.}
For every time step and magnetic quantity, an average was computed along the 
contours, thereby creating time series of the form 
$X(t)\in\left\{\left<B_\text{ver}\right>_\psi(t)\right.$, 
$\left<|B|\right>_\psi(t)$, 
$\left.\left<\gamma_\textsc{lrf}\right>_\psi(t)\right\}$.
Similarly, standard deviations along the contours $\sigma_\psi(t)$ were 
calculated.
These time series (c.f Sect.~\ref{sec:results} and  
Figs.~\ref{fig:1stH0.5}~and~\ref{fig:1stH0.4}) 
show a daily variation of an approximately 
sinusoidal shape. We believe them to be an artefact of SDO's geosynchronous orbital motion. 
For the ranges from $t_\text{start}$ to $t_\text{end}$ given in 
Table~\ref{tab:time}, these time series are least square fitted against 
functions of the form
\begin{align}
X_\text{fit}(t)&=X_0+X_1\sin(2\pi t)+X_2\cos(2\pi t)\notag\\
&=X_0+\underbrace{\sqrt{{X_1}^2+{X_2}^2}}_{X_3}
\cos\Bigg(2\pi t-\underbrace{\arctan\frac{X_1}{X_2}}_{X_4}\Bigg),
\label{eq:fit_t}
\end{align}
whereby $t$ is in days and $t\in\mathbb{N}$ is at noon.
$X_0$ is the value we are interested in and will be henceforth called offset.
It is used instead of a time average $\langle X(t)\rangle_t$ because 
it correctly accounts for missing data (most importantly the gap in the 
afternoon of Oct 17) and that $t_\text{start}$ \& $t_\text{end}$ have a 
different time of day. Here we have
$|\langle X(t)\rangle_t-X_0|<0.5\,\text{G}$ for all $X(t)$ in G and 
$<0.01\degr$ for $X(t)=\left<\gamma_\textsc{lrf}\right>_\psi(t)$.
While $X_1\ \&\ X_2$ are used internally during the fitting process to 
guarantee numeric stability, the results are presented with parameters 
${X_0,\ X_3,\ and\ X_4}$ in Table~\ref{tab:res}. 
$X_3$ and $X_4$ are the amplitude and phase of the orbital artefacts.
Also listed are the standard deviations of the residuals 
${\sigma_t=\sigma\left(X(t)-X_\text{fit}(t)\right)}$ 
and the means of the standard deviations along the contours over the same 
range in time ${\langle\sigma_\psi\rangle_t}$.

\begin{table*} 
    \caption{Timestamps of our spot, year=2012} 
    \label{tab:time}      
    \centering                          
    \begin{tabular}{c | c c c c | c @{} r r}
        \hline\hline                 
        NOAA AR & $t_\text{<E}$ & $t_\text{start}$ & 
        $t_\text{end}$ & $t_{>W}$ & $t_\text{ref}$ & 
        \rule{0pt}{2.2ex}Stonyhurst Lon & Lat\\
        \hline                        
        11591                           & 10.11. 17:24 & 10.13. 19:24 & 10.22.  22:24
        & 10.25. 08:00 & 10.17. 23:59:59Z & \rule{0pt}{2.2ex}$-7$ & +7\\
        \hline                                   
    \end{tabular}
\end{table*}

\begin{table*}  
    \caption{Results: fit parameters and time averages
                }
    \label{tab:res}      
    \centering                          
    \hfill\begin{math}\begin{array}{@{}r r r || r r r r | r | r | r }
    \hline\hline                 
    I_\text{c}/I_\text{qs} & \hspace*{-2.5pt}\Delta h[\text{km}] & 
    \multicolumn{1}{l||}{\text{variable}} & 
    X_0 & X_3 & X_4 [rad] & \sigma_t & 
    \langle\sigma_\psi\rangle_t & \left<Y_3\right>_t &
    \left< d\right>_{\psi,t} [px]
    \rule{0pt}{2.2ex}\\
    \hline                        
    \rule{0pt}{2.4ex} 
    0.50 & 465 & -B_\text{ver}   [\text{G}] & 1693  & 18  &  3.067 & 15  &  81 &  41 & 0.45\\
    0.50 &   0 & -B_\text{ver}   [\text{G}] & 1695  & 17  &  2.981 & 16  & 113 &  97 & 0.59\\
    0.40 & 465 & -B_\text{ver}   [\text{G}] & 1850  & 20  &  3.109 & 19  &  83 &  58 & 0.50\\
    0.40 &   0 & -B_\text{ver}   [\text{G}] & 1849  & 18  &  3.107 & 21  & 114 & 103 & 0.66\\
    0.50 & 465 & |B|             [\text{G}] & 2171  & 16  & -3.120 & 48  & 111 & 102 & 0.97\\
    0.50 &   0 & |B|             [\text{G}] & 2175  & 14  &  3.116 & 47  & 124 & 112 & 1.09\\
    0.40 & 465 & |B|             [\text{G}] & 2265  & 17  & -3.118 & 54  & 117 & 116 & 1.18\\
    0.40 &   0 & |B|             [\text{G}] & 2267  & 16  & -3.082 & 55  & 131 & 126 & 1.31\\
    0.50 & 465 & \gamma_\textsc{lrf} [\degr]& 141.4 & 0.2 &  2.835 & 1.6 & 2.6 & 2.5 & \tablefootmark{a}0.82\\
    0.50 &   0 & \gamma_\textsc{lrf} [\degr]& 141.4 & 0.2 &  2.728 & 1.5 & 2.8 & 2.6 & \tablefootmark{a}0.84\\
    0.40 & 465 & \gamma_\textsc{lrf} [\degr]& 145.0 & 0.2 &  2.968 & 1.6 & 2.5 & 2.3 & 0.77\\
    0.40 &   0 & \gamma_\textsc{lrf} [\degr]& 144.8 & 0.2 &  2.943 & 1.4 & 2.5 & 2.3 & 0.77\\
    \hline                        
    \rule{0pt}{2.4ex} 
    0.53 & 465 & -B_\text{ver}   [\text{G}] & 1639  & 17  &  3.058 & 15  &  82 &  38 & 0.44\\
    \hline                                   
    \end{array}\end{math}\hfill\null\newline
    \addtolength{\arraycolsep}{1.75pt}
    \null\tablefoottext{a}{Excluding five snapshots due to faulty 180\degr-disambiguation: t=10.22. \{12:14,12:36,13:36,14:00,14:12\}}
\end{table*}

\paragraph{Levels of magnetic contours.}
The offsets $X_0$ from the fits to 
$\left<B_\text{ver}\right>_\psi(t)$, $\left<|B|\right>_\psi(t),$ and
$\left<\gamma_\textsc{lrf}\right>_\psi(t)$ for the $0.5\ (0.4)\,I_\text{qs}$ 
contours are then used as contour level 
on the $B_\text{ver}$, $|B|$ and $\gamma_\textsc{lrf}$ maps, respectively.
They are discussed in Sect.~\ref{subsec:contours} and plotted in 
Figs.~\ref{fig:scr}~and~\ref{fig:scr1st0.5_-60}
and the videos.

\paragraph{Distance between contours.}
To quantify how well two contours match we calculated the average distance 
between them $\langle d\rangle_{\psi}$, which we define as the area of 
symmetric difference divided by the length of the intensity contour, $\ell(t)$.
The area of symmetric difference, $\Delta a(t)$, is the area surrounded by either of the contours but not both.
When averaging in time we weighed by the contour length, giving 
\begin{equation}\label{eq:distance}
\langle d\rangle_{\psi,t}={\sum_t \Delta a(t)}\Big/{\sum_t \ell(t)}. 
\end{equation} 
%
These average distances between contours are listed in 
Table~\ref{tab:res} in pixel.
For $\langle d\rangle_{\psi}\ll1\text{pixel}$ only the total ordering 
should be relied upon due to griding and other computational effects.

\newlength{\mySWFheight}\setlength{\mySWFheight}{\textwidth-3.6\baselineskip}
\begin{sidewaysfigure*}
    \centering
    \vspace*{0.5\baselineskip}\includegraphics[height=\mySWFheight]{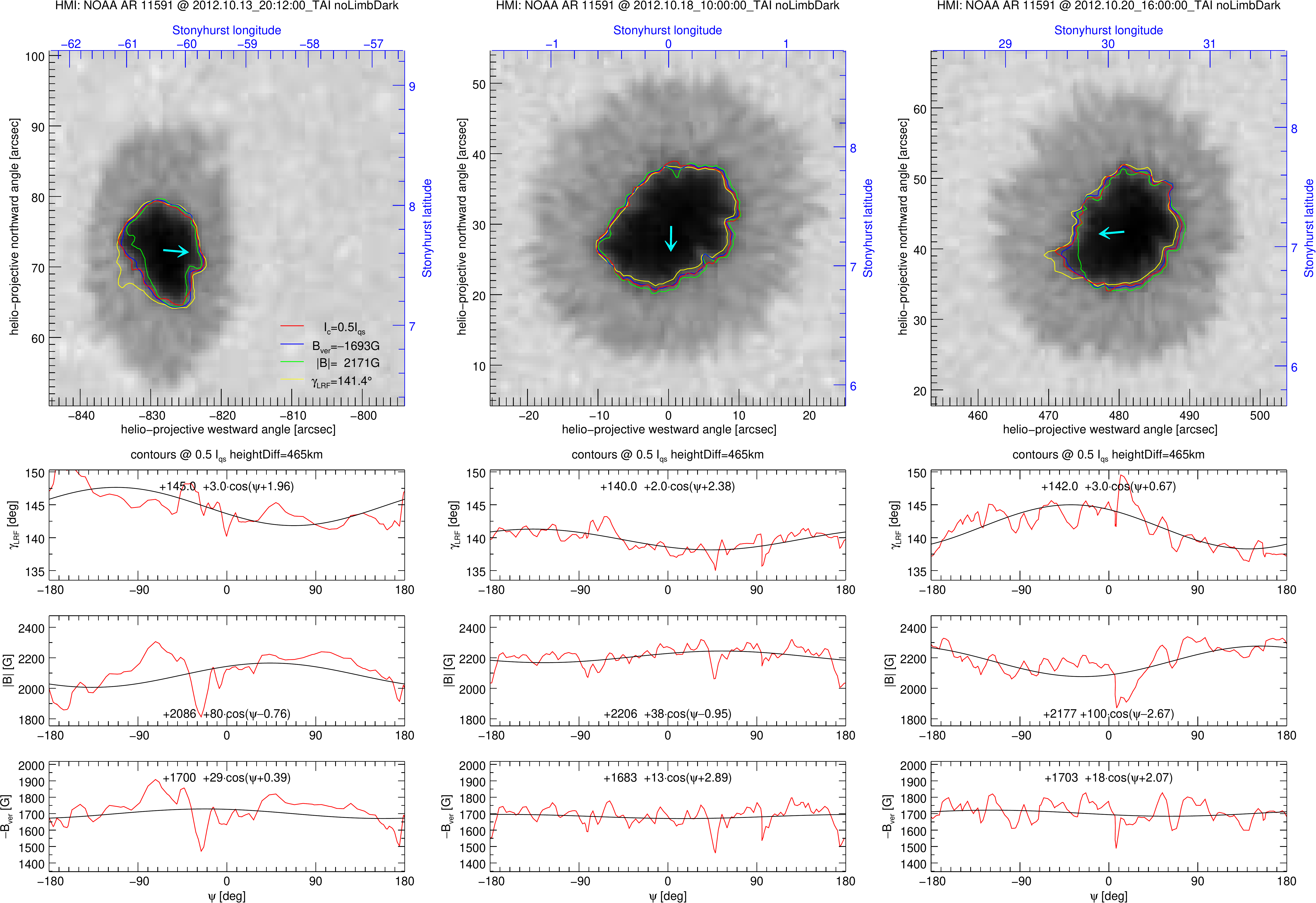}
    \caption[continuum map and contours of NOAA AR 11591]%
    {   NOAA AR 11591, the top panels show continuum 
        intensity maps for longitudes $-60\degr$, $0\degr$ \& $30\degr$.
        The legend in the lower right corner of the top left panel defines the contour levels.
        Different formation heights are accounted for 
        (Eq.~\ref{eq:trans}, $\Delta h=465\,\text{km}$).
        The cyan arrow originates in the centroid of the umbra and points towards disc centre.
        The bottom 3 rows show the magnetic field 
        parameters retrieved along the 
        $I_\text{c}=0.5I_\text{qs}$ contour.
        \hfill\mbox{The temporal evolution is available online at 
            \href{http://no.such.adress/exists}{http://www.aanda.org}}
    }%
    \label{fig:scr}%
\end{sidewaysfigure*}

\paragraph{Fit along each contour.}
For every point along a contour, a reference angle $\psi=\measuredangle(PCD)$ 
is  calculated, whereby $P$ is the point on the contour, 
$C$ is the centroid of the $I_\text{c}=0.5I_\text{qs}$ contour in the CCD frame and 
$D$ is the centre of the solar disc as observed by SDO. 
The angles are calculated on the sphere.
For every time step and every contour, 
$Y(\psi)\in\left\{B_\text{ver}(\psi),\ |B|(\psi),\ 
\gamma_\textsc{lrf}(\psi)\right\}$ 
is least square fitted against functions of the form 
\begin{align}
Y_\text{fit}(\psi)&=Y_0+Y_1\sin(\psi)+Y_2\cos(\psi)\notag\displaybreak[0]\\
&=Y_0+\underbrace{\sqrt{Y_1^2+Y_2^2}}_{Y_3}
\cos\Bigg(\psi-\underbrace{\arctan\frac{Y_1}{Y_2}}_{Y_4}\Bigg).\label{eq:fit_p}
\end{align}
Those fits are plotted in the right panels of the videos 
(cf. Sect.~\ref{subsec:contours} and bottom panels of Figs.~\ref{fig:scr}~and~\ref{fig:scr1st0.5_-60}).
Furthermore the time averages of the fit amplitudes $\left<Y_3(t)\right>_t$ are 
listed in Table~\ref{tab:res}.

\paragraph{Optimal height difference.}
\phantomsection\nobreak\label{par:whatH}\nobreak%
$\Delta h=465\,\text{km}$ was chosen because it minimizes 
the average distance $\langle d\rangle_{\psi,t}$ between the 
$I_\text{c}=0.5I_\text{qs}$ contours after transformation with 
Eq.~\ref{eq:trans} and the $B_\text{ver}$ contours, 
whereby the contour level $X_0$ on the $B_\text{ver}$ map has been derived 
with the fit to $\langle B_\text{ver}\rangle_\psi(t)$ as described above
(Eq.~\ref{eq:fit_t}).
An optimal height difference of ${\Delta h=465\,\text{km}}$ means that the intensity contour at the limb is 
shifted outwards by 
$465\,\text{km}\cdot {r_\sun}/R_\sun\approx0.65\arcsec\approx1.3\text{ pixel}$. 
The difference of the formation heights for continuum and \ion{Fe}{I}~617.3\,nm line core amounts to $\approx$250 km for a typical umbral model atmosphere \citepads[see e.g.][Table 1]{2006SoPh..239...69N}. The fact that the value for $\Delta h$ is larger may be explained with the Wilson depression of the umbra, which typically amounts to 800\,km. The latter causes the $\tau=1$ surface to be strongly inclined relative to horizontal.
Minimizing the standard deviation of $B_\text{ver}$ along the 
$I_\text{c}=0.5I_\text{qs}$ contour ($\langle\sigma_\psi\rangle_t$ column in 
Table~\ref{tab:res}) instead would give an optimal $\Delta h=520\,\text{km}$. 

\section{Results}
\label{sec:results}
Based on the time series of approximately ten days, in which the spot of NOAA AR 11591 has heliocentric angles smaller than 60\degr, we determine the magnetic properties for two distinct contour levels of the continuum intensity. As intensity levels we use  ${I_\text{c}=0.5\ (0.4)\,I_\text{qs}}$. Along each contour, the azimuthal average of $B_{\text {ver}}$, $|B|$ and $\gamma_\textsc{lrf}$ are calculated. 
The respective values  of those averages for $B_{\text {ver}}$ (in blue) and $|B|$ (in black) as well as sinusoidal fit of the orbital variation are displayed in the upper panels of Fig.~\ref{fig:1stH0.5} for ${I_\text{c}=0.5\,I_\text{qs}}$ and of Fig.~\ref{fig:1stH0.4} for ${I_\text{c}=0.4\,I_\text{qs}}$. 
The lower panels show the residuals after subtracting the fit. 

\subsection{Temporal evolution}
The parameters of the sinusoidal fits, offset $X_0$, amplitude $X_3$, and the rms of the corresponding residuals, $\sigma_t$, are given in Table 2  for all considered cases. In addition, they are printed into the plots of Figs.~\ref{fig:1stH0.5} and \ref{fig:1stH0.4}. For the contours at ${I_\text{c}=0.5\,I_\text{qs}}$, 
we find for $\left<B_\text{ver}\right>_\psi(t)$ that $\sigma_t=15$\,G 
is smaller than the orbital amplitude $X_3=18$\,G, 
with an offset of $X_0=1693$\,G. 
For the contours of ${I_\text{c}=0.4\,I_\text{qs}}$, $\sigma_t=19$\,G 
is also smaller than $X_3=20$\,G 
with an $X_0=1850$\,G. For the residuals of $B_\text{ver}$ no long-term trend is noticeable. 

In contrast, the residuals of $\left<|B|\right>_\psi(t)$ amount to $\sigma_t=48$\,G 
which is larger than the amplitudes of the sinusoidal fit (16\,G), and it shows a long-term variation. 
Since $\gamma_\textsc{lrf}$ is dependent on $B_\text{ver}$ and $|B|$, it has a long-term variation which compensates for that of $|B|$ (not shown). The offsets $X_0$ for $\left<|B|\right>_\psi(t)$ and 
$\left<\gamma_\textsc{lrf}\right>_\psi(t)$ are $2171\,\text{G}$ and $141.4\degr$ respectively at the contours with ${I_\text{c}=0.5\,I_\text{qs}}$. 

The fact that the residuals $\sigma_t(\left<B_\text{ver}\right>_\psi(t))$ are smaller than $\sigma_t(\left<|B|\right>_\psi(t))$ is remarkable, but is even more remarkable if one considers that the gradient of $B_\text{ver}$ perpendicular to the contour is larger than that of $|B|$. This can be inferred from Table~\ref{tab:res}: The difference of the $\left<B_\text{ver}\right>_\psi(t)$ offset, $X_0$, 
between the two different intensities amounts to 157\,G 
while that of $\left<|B|\right>_\psi(t)$ is only 94\,G. Hence, a small shift of the contour implies a larger deviation in $B_\text{ver}$ than in $|B|$. 
Therefore, our result of a smaller deviation in $B_\text{ver}$ relative to $|B|$ gives further evidence that $\left<B_\text{ver}\right>_\psi(t)$ can be considered constant in time.

\begin{figure}
    \includegraphics[width=\columnwidth]{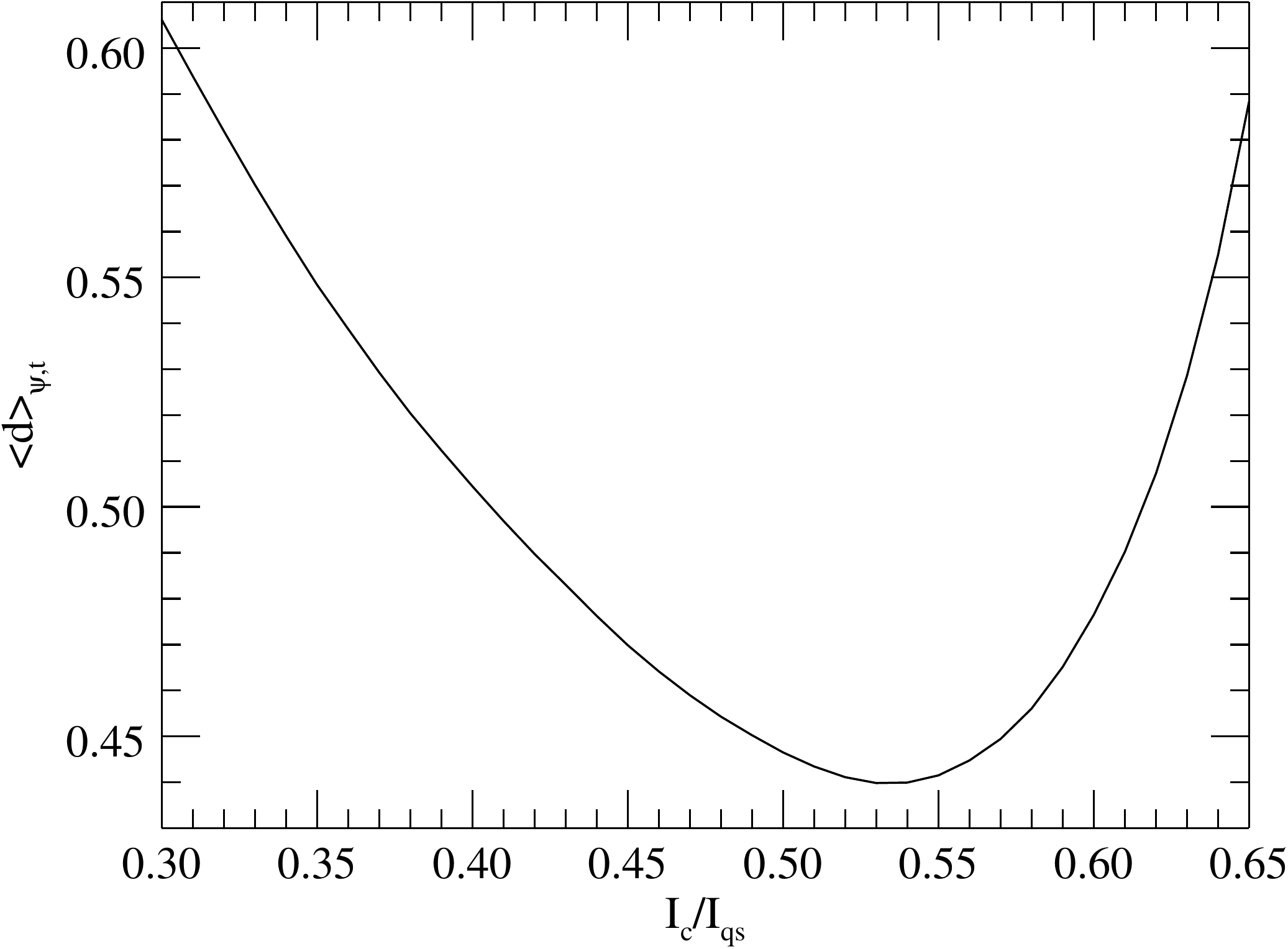} 
    \caption[dVsR]
    {Average distance, $\left<d\right>_{\psi,t}$, between contours of varying intensity and $B_\text{ver}$. The contour level of $B_\text{ver}$ is determined by fitting Eq.~(\ref{eq:fit_t}). $\left<d\right>_{\psi,t}$ has a minimum for $I_\text{c}=0.53I_\text{qs}$ corresponding to $-B_\text{ver}=1639$\,G. }
    \label{fig:distance}
\end{figure}

\subsection{Contours}
\label{subsec:contours}
Using the offsets $X_0$ from the fits in Table~\ref{tab:res} with $I_\text{c}=0.5I_\text{qs}$ and $\Delta h=465\,\text{km}$, 
the upper panels 
of Fig.~\ref{fig:scr} overplot 
the contours of intensity $I_\text{c}=0.5I_\text{qs}$ 
(red), 
$|B|=2171\,\text{G}$ (green), 
$-B_\text{ver}=1693\,\text{G}$ (blue), 
and $\gamma_\textsc{lrf}=141.4\degr$ (yellow). 
The background images consist of 
100x100\;pixel cutouts of grey-scale intensity maps with a minimum (maximum) of  ${I_\text{c}=0.1\ (1.2)\ I_\text{qs}}$. 
A close inspection of the figure shows that the $B_\text{ver}$ contour matches best with the intensity contour.
The cyan arrow originates in 
the centroid of the umbra and points towards disc centre. The centroid is 
determined by the $I_\text{c}=0.5I_\text{qs}$ contour and 
is derived using 
CCD coordinates. 

The three bottom rows of 
panels of Fig.~\ref{fig:scr} show the magnetic field quantities along the $I_\text{c}=0.5I_\text{qs}$ contour as well as their sinusoidal fits in black. 
The azimuth is determined relative to the centroid and the direction towards disc centre, 
which corresponds to $\psi=0\degr$ and runs counter-clockwise.

To quantify the azimuthal variation of the magnetic parameters, Table 2 gives the time average of the standard deviations along the contours, $\langle\sigma_\psi\rangle_t$. 
Again $\langle\sigma_\psi\rangle_t$ is smaller for $B_\text{ver}$ (81\,G) than for $|B|$ (111\,G). 
As before, the small value for $B_\text{ver}$ is remarkable, since its gradient perpendicular to the contour is larger than for $|B|$.
%
The lower panels demonstrate that the azimuthal variations are smallest for $B_\text{ver}$. Again, we note that this is remarkable considering the fact that the gradient of $B_\text{ver}$ perpendicular to the contour is larger than the gradient of $|B|$.

A video of the temporal evolution of those contours during the disc passage of the spot is available at \href{http://no.such.adress/exists}{http://www.aanda.org}. 
This animation demonstrates that an iso-contour of $B_\text{ver}=-1693$\,G coincides nicely with the intensity contour at $0.5I_\text{qs}$. 
This animation also demonstrates that contours of $|B|$ and $\gamma_\textsc{lrf}$ do not coincide. 

To quantify the match or mismatch of two contours, we have introduced the average distance between two sets of contours, $\left<d\right>_{\psi,t}$ (cf. Eq.~\ref{eq:distance}). It is given in the last column of Table~\ref{tab:res}.
$\left<d\right>_{\psi,t}$ is smallest for the $B_\text{ver}$ contours with $\Delta h=465\,\mathrm{km}$ (see the first and final row of Table~\ref{tab:res}).

In Fig.~\ref{fig:distance} the average distance is plotted for intensities changing from 0.30 to 0.65. The corresponding contour levels for $B_\text{ver}$ are calculated as described in Sect.~\ref{sec:methods} (fit to Eq.~\ref{eq:fit_t}). 
The best match, $\left<d\right>_{\psi,t} =0.44 $, is found for $I=0.53I_\text{qs}$ with 
$-B_\text{ver}=1639$\,G ($X_3=17$\,G, $\sigma_t=15\,$G, 
and $\langle\sigma_\psi\rangle_t=82\,$G). 
Distances for $|B|$ and $\gamma_\textsc{lrf}$ are in all cases larger and not plotted. Hence, by minimizing the distance, 
$-B_\text{ver}=1639$\,G results as the value that defines the umbral boundary at $I=0.53I_\text{qs}$. 
This is additional proof that our chosen value of $I=0.5I_\text{qs}$ is very close to the optimum value.

\begin{figure}
\centering\includegraphics[height=\textwidth-3.6\baselineskip]{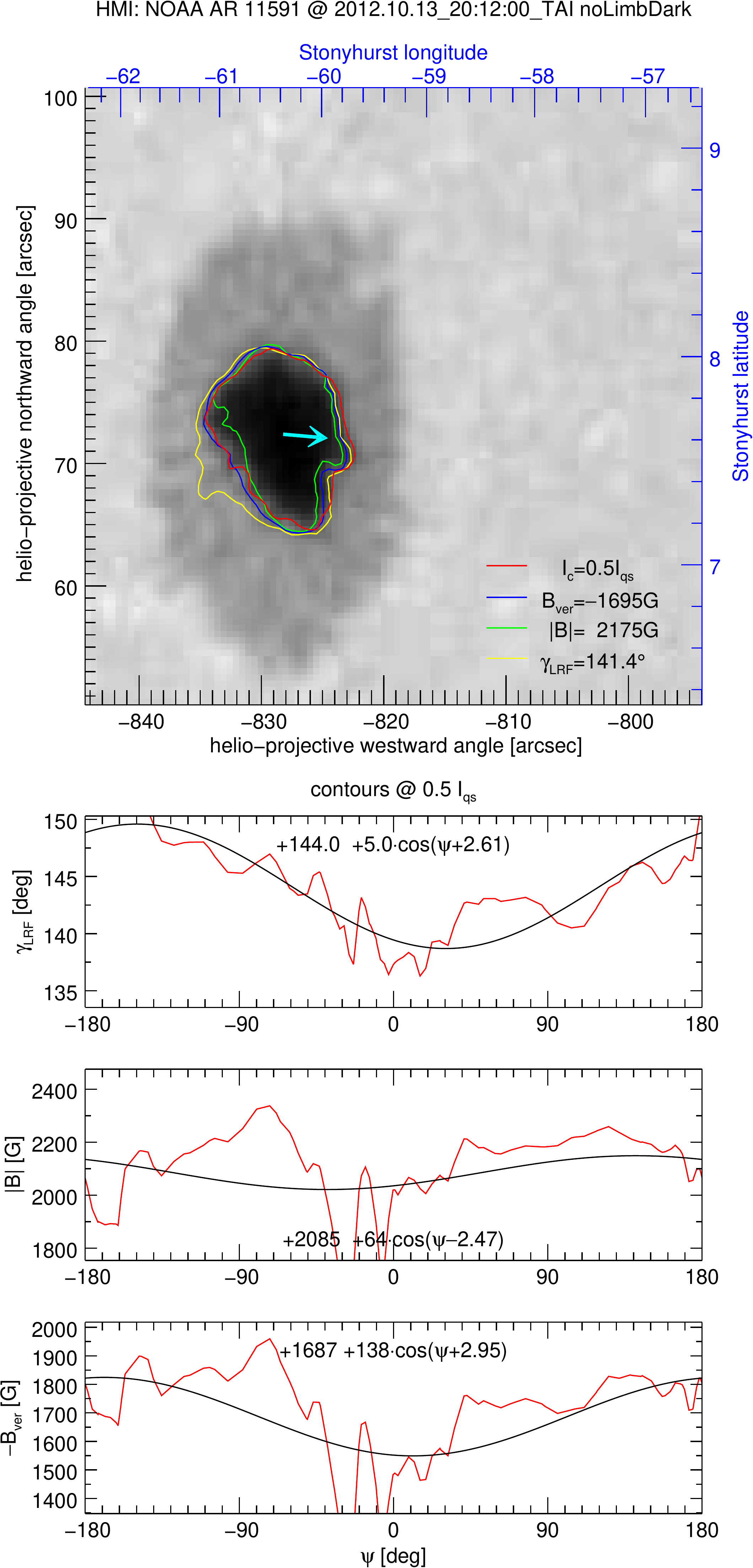}
    \caption{as Fig.~\ref{fig:scr}, left column, 
        but without compensation for different formation heights ($\Delta h=0$).
    \newline\null\hfill The temporal evolution is available online at
        \href{http://no.such.adress/exists}{http://www.aanda.org}}%
    \label{fig:scr1st0.5_-60}
\end{figure}

\subsection{Effect of neglecting formation heights compensations}\label{sec:whyH}
For the results presented so far, we corrected for the projection effects due to different formation heights of continuum and line. As discussed in the end Sect.~\ref{par:whatH} we assume a height difference of $\Delta h=465\,\text{km}$. Table 2 also gives the results for the case in which these projection effects are not considered, i.e. $\Delta h=0\,\text{km}$. As a general trend, it is seen that the values for $X_0$, $X_3$, and $\sigma_t$ change only marginally. A plot like in Fig.~1 with $\Delta h=0\,\text{km}$ looks almost identical (not shown). 

However, $\langle\sigma_\psi\rangle_t$ and $\left< d \right>_{\psi,t}$ increase significantly. For example, for $B_\text{ver}$ at $I=0.5I_\text{qs}$, $\langle\sigma_\psi\rangle_t$ and $\left< d \right>_{\psi,t}$ 
increase by more than 30\% from 81 to 113\,G, and from 0.45 to 0.59\,pixel, respectively. 
This is illustrated in Fig.~\ref{fig:scr1st0.5_-60}, which shows the same snapshot as in the left column of 
Fig.~\ref{fig:scr}, 
with the only difference that $\Delta h=0\,\text{km}$. In this case, the heliocentric angle is 60\degr. It is seen that the magnetic contours are shifted relative to the intensity, which results in an increase of $\left< d \right>_{\psi,t}$, and the variation of $B_\text{ver}$ along the contour (bottom panels) are larger for $\Delta h=0\,\text{km}$.
This can also be seen in the corresponding video of the disc passage of the spot, which is available at \href{http://no.such.adress/exists}{http://www.aanda.org}

\section{Conclusion}

Investigating the physical properties along the umbra-penumbral boundary of a stable sunspot for a time span of approximately ten, 
we find three main results:
\begin{enumerate}
\item $B_\text{ver}$ averaged along the $I=0.5I_\text{qs}$ contour is nearly constant in time.
\item Contours of intensity and of $B_\text{ver}$ match at the umbral boundary. The best match is obtained for  $I=0.53I_\text{qs}$ and 
$|B_\text{ver}|=1639$\,G.
\item Projection effects due to different formations height of the spectral line and continuum need to be considered. If not, variation of $B_\text{ver}$ along the contour increases significantly.
\end{enumerate}
These results are obtained by analysing 1063 consecutive SDO/HMI data sets (with a time step of 12 min) of the first disc passage of NOAA AR 11591.

Using ${I_\text{c}=0.5\ I_\text{qs}}$ to define the umbral boundary, we obtain 
$|B_\text{ver}|=1693\,\text{G}\pm 15$ (1\,$\sigma_t$-error). 
\citetads{2018A&A...611L...4J} used Hinode/SP data to find ${|B_\text{ver}|=1867_{-16}^{+18}G}$ (99\%-error) at $I_\text{c}=0.5\ I_\text{qs}$. 

The values for $|B_\text{ver}|$ differ by some 175\,G. In general, a difference is expected due to differences in the experimental setup and analysis methods. \citetads{2017ApJ...851..111S} investigates the differences between HMI and SP vector magnetograms and obtained comparable differences. 
He concludes that the filling factor followed by 
spatial and spectral resolution are the main source. 
At the umbral boundary the filling factor is 1, and causes therefore no differences. The other 
effects are particularly strong at the sharp boundary between umbra and penumbra, where the intensity gradient is large.

Hence, these investigations provide evidence that $|B_\text{ver}|$ is constant for a statistical sample of sunspots 
as well as during the evolution of one stable spot, 
thereby supporting the {Jur{\v c}{\'a}k} criterion. 
%


\begin{acknowledgements}
We wish to thank Jan {Jur{\v c}{\'a}k}, Juan Manuel Borrero and the anonymous reviewer for 
valuable discussions, Xudong Sun for making available the vector transformation 
routine as well as Hanna Strecker and various members of the 
\href{http://www.idlcoyote.com/comp.lang.idl-pvwave/}
{IDL user group} for help with IDL and 
\href{https://tex.stackexchange.com}{tex.sx} users for help with LaTeX.

The data used is courtesy of NASA/SDO and the HMI science team 
(see e.g. \citealtads{1994SoPh..155..235M}; 
\citealtads{2009SoPh..260...83L}; 
\citealtads{2011SoPh..273..267B}; 
\citealtads{2012SoPh..275....3P}; 
\citealtads{2012SoPh..275..229S}; 
\citealtads{2014SoPh..289.3483H}). 

This research has made use of NASA's 
\href{http://adsabs.harvard.edu/abstract_service.html}%
{Astrophysics Data System.}
\end{acknowledgements}

\bibliographystyle{aa}
\bibliography{schmassmannAA,schmassmannMSc}

\end{document}